\documentclass[prx,twocolumn,amsmath,amssymb]{revtex4-2}

\usepackage[dvipdfmx]{xcolor}
\usepackage{bm}
\usepackage{dcolumn}
\usepackage[pdftex]{graphicx}    

\usepackage{slashed}
\usepackage{amsmath}\usepackage{accents}
\usepackage{physics}

\newcommand{\cin}[1]{{\color{black}#1}}

\newcommand{\1}{\mbox{1}\hspace{-0.25em}\mbox{l}}

\begin{document}

\title{
Real-space representation of the second Chern number
}

\author{Tsuyoshi Shiina, Fumina Hamano, and Takahiro Fukui}
\affiliation{Department of Physics, Ibaraki University, Mito 310-8512, Japan}

\date{\today}

\begin{abstract}
We extend Kitaev's real-space formulation of the first Chern number to the second Chern number and establish a computational framework for its evaluation. To test its validity, we apply the derived formula to the disordered Wilson-Dirac model and analyze its ability to capture topological properties in the presence of disorder. Our results demonstrate that the real-space approach provides a viable method for characterizing higher-dimensional topological phases beyond momentum-space formulations.
\end{abstract}

\pacs{
}

\maketitle

\section{Introduction}

Topological invariants play a crucial role in characterizing phases of matter 
beyond the conventional symmetry-breaking paradigm \cite{Hasan:2010fk,Qi:2011kx}. 
Among them, the Chern number provides a fundamental classification 
of topological insulators and superconductors in even-dimensional systems \cite{Thouless:1982uq,kohmoto:85,Qi:2008aa,Schnyder:2008aa,Thonhauser:2005aa,Ceresoli:2006aa,doi:10.1143/JPSJ.77.123705,Price:2015ab,Price:2016aa,Fukui:2016aa,Fukui:2017aa,Lohse_2018,Zilberberg_2018,Mochol-Grzelak_2019,Zhang:2023ab}. 
While the first Chern number is well understood and widely utilized in the context of quantum Hall systems 
and time-reversal symmetry-breaking insulators, 
the second Chern number has gained attention in higher-dimensional generalizations, 
such as four-dimensional quantum Hall systems \cite{Price:2015ab,Fukui:2016aa,Lohse_2018,Zilberberg_2018}, three-dimensional pumps
\cite{Qi:2008aa,Fukui:2017aa,Lohse_2018,Zilberberg_2018},
and the Berry curvature description of generic time-reversal invariant systems \cite{Hatsugai_2010}, etc.

The calculation of topological invariants in real space becomes essential for studying disordered and inhomogeneous systems, where momentum space methods fail. Several real-space methods for computing first Chern numbers have been developed. One approach is to use the twisted boundary condition \cite{Niu:1985fr}, where the twist angles serve as momenta. Another is the so-called Chern marker method \cite{PhysRevB.84.241106}, in which derivatives with respect to momenta are represented by position operators.
Kitaev's method \cite{Kitaev:2006yg} provides a real-space calculation of the first Chern number by utilizing projectors onto specific subsets of lattice sites, rather than employing position operators. In this approach, 
the entire lattice is divided into three disjoint regions, and a central role is played by the projectors onto them
rather than the position operators.
This method has been successfully applied to various models, 
which do not allow simple momentum space calculations
\cite{PhysRevB.108.085114,PhysRevB.110.245405}.
Kitaev's approach has the advantage of being manifestly topological, as it is invariant 
under the deformation of artificially divided regions.

The rest of this paper is organized as follows. In Sec. \ref{s:berry}, several useful formulas associated 
with the spectral projector are summarized. 
In Sec. \ref{s:ch1}, we review Kitaev's real-space formulation of the first Chern number and its key properties. 
We also check its effectiveness by applying it to the Wilson-Dirac model in two dimensions. 
In Sec. \ref{s:ch2}, we present the extension to the second Chern number and discuss its theoretical foundation. 
We also apply our formula to the Wilson-Dirac model in four dimensions and analyze the numerical results. 
Finally, in Sec. \ref{s:sum}, we summarize our findings and outline future research directions.

\section{Berry connection and curvature}\label{s:berry}

In this section, we summarize several useful formulas associated with the spectral projectors in order to represent
the Chern numbers in higher dimensions \cite{Hatsugai_2010}.

We begin with the Bloch Hamiltonian ${\cal H}(q)$ for systems with translational invariance on lattices in $d=2, 4$, or more generally, $2n$ dimensions, where the Schr\"odinger equation is given by  
\begin{alignat}1
{\cal H}(q)\psi_{n}(q)=\varepsilon_n(q)\psi_n(q).
\end{alignat}  
We assume a gapped ground state composed of $M$-multiplet wave functions, 
$\psi(q) \equiv (\psi_{1},\psi_{2},\cdots,\psi_{M})$.  
Using this, we define the Berry connection one-form and the Berry curvature two-form as  
\begin{alignat}1
A=A_\mu dq_\mu,\quad F=dA+A^2=\frac{1}{2}F_{\mu\nu}dq_\mu dq_\nu,
\end{alignat}  
where $A_\mu=\psi^\dagger\partial_\mu\psi$ and  
$
F_{\mu\nu}=(\partial_\mu\psi^\dagger \partial_\nu\psi-\partial_\nu\psi^\dagger \partial_\mu\psi)+[\psi^\dagger \partial_\mu\psi,\psi^\dagger \partial_\nu\psi].
$
Throughout this paper, derivatives are taken with respect to momenta, i.e., $\partial_\mu\equiv\partial_{q_\mu}$.  

The Berry curvature can be expressed in terms of the spectral projector,  
\begin{alignat}1
P(q)=\psi(q)\psi^\dagger(q),
\end{alignat}  
and its exterior derivative with respect to $q_\mu$,  
\begin{alignat}1
dP=\partial_{\mu} P \, dq_\mu.
\end{alignat}  
First, note that  
\begin{alignat}1
\psi^\dagger dP =\psi^\dagger (d\psi\psi^\dagger+\psi d\psi^\dagger)
= d\psi^\dagger(1-\psi\psi^\dagger)\equiv d\psi^\dagger \bar P,
\label{PsidP}
\end{alignat}  
where $\bar P\equiv 1-P$ is the complement of $P$. Likewise, we have  
\begin{alignat}1
dP\psi=\bar P d\psi.
\label{dPPsi}
\end{alignat}  
On the other hand, since $P^2=P$, we have $dPP + PdP = dP$, which naturally leads to  
\begin{alignat}1
dP\bar P=PdP, \quad dPP=\bar P dP.
\end{alignat}  
Thus, we obtain  
\begin{alignat}1
[(dP)^2,P]=0, \quad [(dP)^2,\bar P]=0.
\end{alignat}  
Using these, together with Eqs.~(\ref{PsidP}) and (\ref{dPPsi}), we find  
\begin{alignat}1
\psi^\dagger (dP)^2\psi &= d\psi^\dagger\bar P d\psi = d\psi^\dagger(1-\psi\psi^\dagger)d\psi
\nonumber\\
&= d\psi^\dagger d\psi + \psi^\dagger d\psi\psi^\dagger d\psi = F,
\end{alignat}  
which leads to  
\begin{alignat}1
F^2 &= \psi^\dagger (dP)^2\psi \, \psi^\dagger (dP)^2\psi= \psi^\dagger (dP)^2 P (dP)^2\psi
\nonumber\\
&= \psi^\dagger P (dP)^4\psi.
\end{alignat}  
Applying this repeatedly yields $F^n = \psi^\dagger P(dP)^{2n}\psi$.  
This expression is useful for representing the Chern numbers in terms of the projector $P$.  
Namely, the $n$th Chern number is given by  
\begin{alignat}1
c_n& = \frac{1}{n!} \left(\frac{i}{2\pi}\right)^n \int \tr F^n
= \frac{1}{n!} \left(\frac{i}{2\pi}\right)^n \int \tr P(dP)^{2n}.
\label{GenChe}
\end{alignat}  
Since $P$ allows a real-space representation, the latter formula serves as a starting point for  
the real-space formulation of the Chern numbers.  

\section{First Chern number}\label{s:ch1}

We review the real-space representation of the first Chern number introduced by Kitaev \cite{Kitaev:2006yg} and establish our notation. This approach can be naturally extended to the second Chern number.

\subsection{The Kitaev formula for the first Chern number}

Let us explore how the first Chern number, originally defined in momentum space, can be reformulated 
in terms of a real-space representation on a lattice. From Eq. (\ref{GenChe}), 
the first Chern number is given by
\begin{alignat}1
c_1
=\frac{i}{2\pi}\int d^2q\,\epsilon^{\mu\nu}\tr P\partial_\mu P\partial_\nu P.
\label{1Che}
\end{alignat}

Let $P_{in,jm}$ be the real-space representation of the spectral projector, 
where $i,j$ represent sites in two dimensions,
i.e., $i=(i_1,i_2)$, and $n,m$ represent the internal degrees of freedom in the unit cell, which give rise to
the non-Abelian nature of the Berry connection.
In what follows, the subscripts $n,m$ will be suppressed for simplicity.
Let $(X_\mu)_{ij}\equiv i_\mu \delta_{ij}$ be the site position operator.
Then, the key matrix for the real-space representation of the first Chern number
(\ref{1Che}) is the following products of the matrices $P$ and $X_\mu$, 
$\left(\epsilon^{\mu\nu}P[X_\mu,P][X_\nu,P]\right)$. 
Its diagonal elements do not depend on the sites
if the system has translational invariance, $P_{jl}=P_{j-l}$, 
\begin{alignat}1
\big(&\epsilon^{\mu\nu}P[X_\mu,P][X_\nu,P]\big)_{ii}
\nonumber\\
&=\sum_{j,k}\epsilon^{\mu\nu}P_{i-j}(j_\mu-k_\mu)P_{j-k}(k_\nu-i_\nu)P_{k-i}
\nonumber\\
&=\sum_{j,k}\epsilon^{\mu\nu}P_{-j}(j_\mu-k_\mu)P_{j-k}(k_\nu)P_{k},
\end{alignat}
where in the second line we have replaced $j\rightarrow j+i$ and $k\rightarrow k+i$. 
This indeed shows that the left-hand side of the above
is independent of $i$.
Translational symmetry also allows  the Fourier transformation
\begin{alignat}1
P_{i-j}=\int_{-\pi}^\pi\frac{d^2q}{(2\pi)^2}e^{iq(i-j)}P(q).
\end{alignat} 
Then,  the above matrix is transformed into 
\begin{alignat}1
\left(\epsilon^{\mu\nu}P[X_\mu,P][X_\nu,P]\right)_{ii}
&=\left(\frac{i}{2\pi}\right)^2\int_{-\pi}^\pi d^2q
\epsilon^{\mu\nu}
P\partial_{\mu} P\partial_{\nu}P.
\label{FouPos_1}
\end{alignat}
Thus we have
\begin{alignat}1
c_1=\frac{2\pi}{i}\epsilon^{\mu\nu}\tr P[X_\mu,P][X_\nu,P]_{ii},
\label{1CheMar}
\end{alignat}
where the trace is taken only over the internal degrees of freedom, not over the lattice site $i$.
Note again that the above expression does not depend on the site index $i$.
This formula forms the basis of the Chern marker \cite{Prodan_2010,PhysRevB.84.241106}.

\begin{figure}[htb]
\begin{center}
\begin{tabular}{c}
\includegraphics[width=0.9\linewidth]{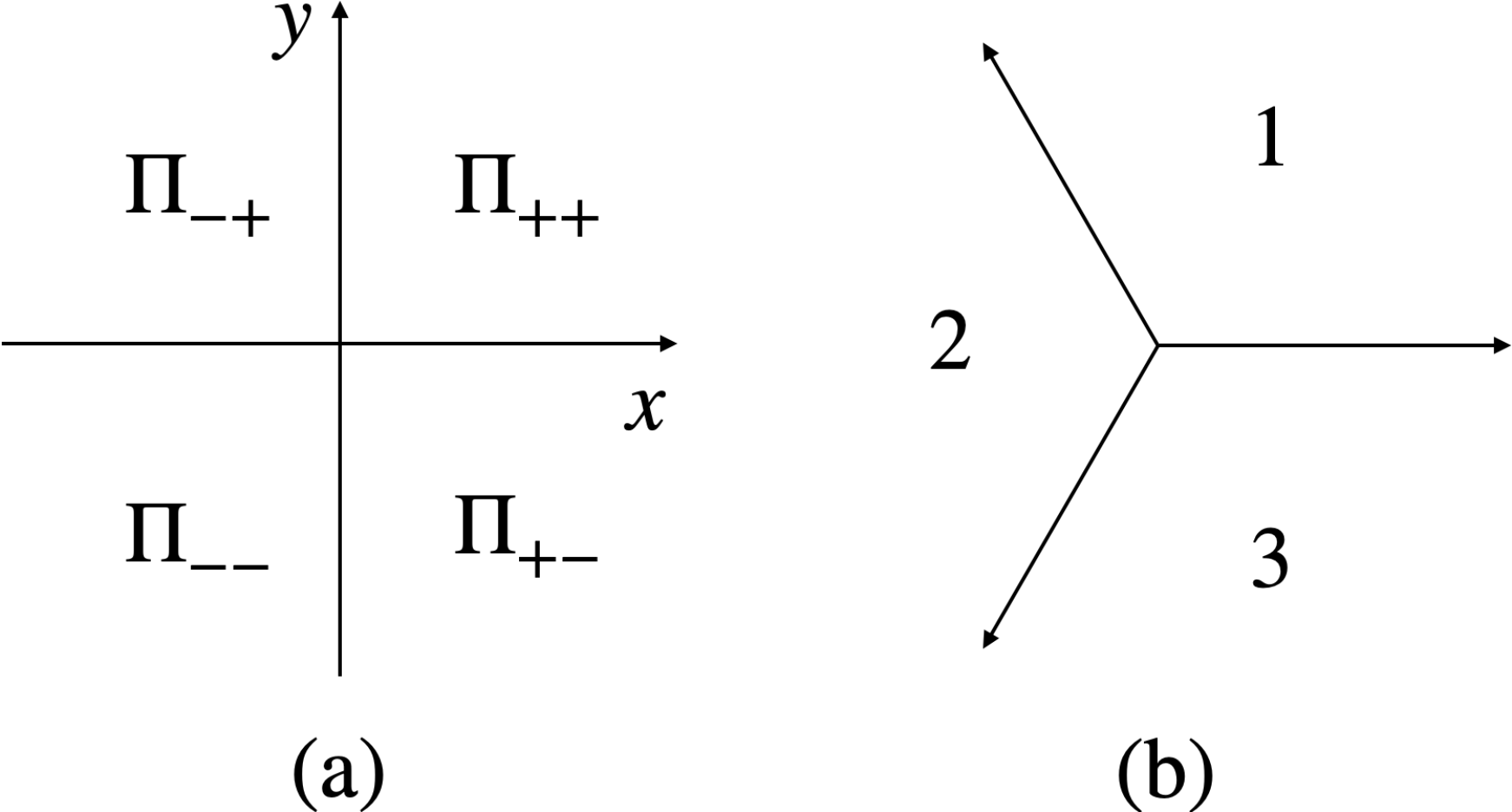}
\end{tabular}
\caption{
Regions in the two-dimensional lattice. (a) With respect to 
$\Pi_x=\Pi_{++}+\Pi_{+-}$ and $\Pi_y=\Pi_{++}+\Pi_{-+}$, 
and (b) with respect to a generic partition.
}
\label{f:ch1}
\end{center}
\end{figure}

Equation (\ref{1CheMar}) can be rewritten into a manifestly topological form \cite{Kitaev:2006yg}.
According to Kitaev, let us introduce the lattice projector $\Pi_\mu$ defined by
\begin{alignat}1
\Pi_\mu=\left\{\begin{array}{ll}\delta_{ij}\quad& (i_\mu\ge 0)\\0&(i_\mu<0)\end{array}\right. =\theta_{i_\mu}\delta_{ij},
\label{LatPro}
\end{alignat}
where $\theta_{i_\mu}=1$ for $i_\mu\ge0$ and $=0$ otherwise.
Then, we have
\begin{alignat}1
\epsilon^{\mu\nu}\tr P[X_\mu,P][X_\nu,P]_{ii}
&=\Tr \epsilon^{\mu\nu} P[\Pi_\mu,P][\Pi_\nu, P]
\nonumber\\
&=\Tr \epsilon^{\mu\nu} P\Pi_\mu P\Pi_\nu P ,
\label{ProPos_1}
\end{alignat}
where Tr stands for the trace over the sites $i$ as well as the internal degrees of freedom $n$, 
$\Tr A=\sum_i\tr A_{ii}=\sum_{i,n}A_{in,in}$. See Refs. \cite{Kitaev:2006yg,PhysRevB.110.045437}.
Hence, we reach
\begin{alignat}1
c_1&=\frac{2\pi}{i}\Tr \epsilon^{\mu\nu} P\Pi_\mu P\Pi_\nu P
\equiv \frac{2\pi}{i}\nu_1(P,\Pi_x,\Pi_y).
\label{1Che1}
\end{alignat} 
Here, note that  $\Pi_x$ and $\Pi_y$ have an overlap region $\Pi_x\Pi_y=\Pi_{++}$, 
where $\Pi_{++}$ stands for the projector onto $j_x\ge0$ and $j_y\ge0$ (see Fig. \ref{f:ch1}).
This overlap region does not contribute to Eq. (\ref{1Che1}) because of the antisymmetrization in the indices $\mu,\nu$.
To remove such a contribution, there are two possibilities:
One is to replace $\Pi_y$ with $\Pi_y(1-\Pi_x)$, and the other is to replace $\Pi_x$ with $\Pi_x(1-\Pi_y)$.
Namely, one contribution 
comes through $\nu_1(P,\Pi_{++}+\Pi_{+-},\Pi_{-+})=\nu_1(P,\Pi_{++},\Pi_{-+})$ and the other comes through
 $\nu_1(P,\Pi_{+-},\Pi_{++}+\Pi_{-+})=\nu_1(P,\Pi_{+-},\Pi_{++})$, each contributing independently,
\begin{alignat}1
\nu_1(P,\Pi_x,\Pi_y)&=\nu_1(P,\Pi_{++},\Pi_{-+})+\nu_1(P,\Pi_{+-},\Pi_{++}).
\label{1Che2}
\end{alignat}
This shows that \(c_1\) is composed of two terms, which are equal, as will be shown below.

Let us divide the two-dimensional lattice into three generic regions labeled $A=1,2,3$, as shown in Fig. \ref{f:ch1}.
The corresponding projectors are defined by $\Pi_A$. 
Then, $\sum_A\Pi_A=1$ and $\Pi_A\Pi_B=\delta_{AB}\Pi_A$ hold.
For such a partition of the lattice, we have the following identity:
\begin{alignat}1
\Tr \epsilon^{ABC}P\Pi_A P\Pi_B P\Pi_C
&=\Tr \epsilon^{AB}P\Pi_A P\Pi_BP
\nonumber\\
&=\nu_1(P,\Pi_1,\Pi_2),
\label{ThrReg}
\end{alignat}
where we have substituted $\Pi_3=1-\Pi_1-\Pi_2$ into the left-hand side of the upper line,
and therefore in the right-hand side, $A,B$ take only 1 and 2.
The two terms in Eq. (\ref{1Che2}) correspond to specific choices of the regions 1 and 2. 
What is important here is that $\nu_1(P,\Pi_1,\Pi_2)$ is invariant under continuous changes of the regions 1, 2, and 3,
which are realized by successively moving one site  from one region into another region 
\cite{Kitaev:2006yg,PhysRevB.110.045437} (see also the proof in four dimensions in Sec. \ref{s:ch2}).
Thus, we have
\begin{alignat}1
c_1&=\frac{2\pi}{ i}2\nu_1(P,\Pi_1,\Pi_2)
\nonumber\\
&=2\frac{2\pi}{ i}\Tr \epsilon^{ABC}P\Pi_A P\Pi_BP\Pi_C.
\end{alignat}
This is the real-space representation derived by Kitaev \cite{Kitaev:2006yg}.
Here, the deformation invariance of $\nu_1$ is due to the following conservation law:
\begin{alignat}1
\sum_{i_3}h_{i_1i_2i_3}=0,
\end{alignat}
where $h$ is defined by
\begin{alignat}1
h_{i_1i_2i_3}&\equiv \epsilon^{abc}\tr P_{i_ai_b}P_{i_bi_c}P_{i_ci_a}.
\end{alignat}
In Sec. \ref{s:ch2}, we provide a proof of the deformation invariance in the case of four dimensions.


\subsection{Numerical calculation for the Wilson-Dirac model}\label{s:2dwd}

In principle, the Kitaev formula is strictly valid only for infinite lattices. When applied to finite systems 
in  numerical computations, Chern numbers vanish
(see the discussions in \cite{PhysRevB.110.045437}).
[In this paper, we have argued that (i) the projector $\Pi_\mu$ produces artificial edge states around 
the intentional jump at $j_\mu = 0$, which are responsible for the nontrivial topological numbers, whereas 
(ii) periodic boundary conditions yield fictitious edge states around the unintentional jump between 
$\Pi_\mu = 0$ at $j_\mu = -N$ and $\Pi_\mu = 1$ at $j_\mu = N-1$, which cancel the topological number 
created around $j_\mu = 0$
.]
To address this issue, Kitaev proposed a
truncation scheme. Specifically, we introduce the following truncation: 
The lattice sites in each direction $\mu$ are assumed to be $-N\le i_\mu\le N-1$,
with the periodic boundary condition imposed.
\cin{Thus, if the system size is specified by $N$, each direction contains $2N$ sites.}
Then, introducing the 
truncation projector $\Pi^{(L)}$ defined by $\Pi^{(L)}\equiv 1$ for $-L\le i_\mu\le L-1$ and $=0$ otherwise,
the Tr operation is modified such that 
\begin{alignat}1
c_1&=2\frac{2\pi}{ i}\Tr \epsilon^{ABC}\Pi^{(L)}P\Pi_A P\Pi_BP\Pi_C.
\label{c1Tru}
\end{alignat}
In what follows, we label the numerical data by the lattice size as well as the truncation size, denoted as $(N,L)$.

\begin{figure}[htb]
\begin{center}
\begin{tabular}{c}
\includegraphics[width=0.95\linewidth]{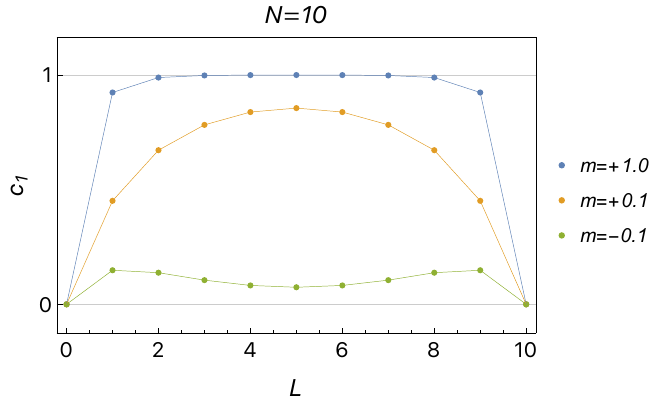}
\end{tabular}
\caption{
The truncation size $L$ dependence of the first Chern number $c_1$ 
computed for the Wilson-Dirac model using the Kitaev formula \cin{(\ref{c1Tru}), 
where we have used $\Pi_1=\Pi_{++}$, $\Pi_2=\Pi_{-+}$, and $\Pi_3=\Pi_{--}+\Pi_{+-}$ 
(see Fig. \ref{f:ch1}). } 
The system size is fixed at $N=10$ (total sites are  $(2N)^2=20^2$), 
and results are shown for  $m=1$, $0.1$, and $-0.1$, which yield
$c_1=1$, $1$ and $0$, respectively. The other parameters are set $t=b=1$.
}
\label{f:1ch_L}
\end{center}
\end{figure}

Now, we investigate the Kitaev formula by applying it to the Wilson-Dirac model in $2n$ dimensions ($n=1$ in this section and $n=2$ in the next section).
To this end, we introduce several notations: Let $j=(j_1,j_2,\cdots,j_{2n})$ denote the lattice site in 
$2n$ dimensions, and let $\hat\mu$ 
be the unit vector in the $\mu$ direction. Then, the Dirac Hamiltonian
on the lattice with the Wilson term is given by 
\begin{alignat}1
H=&-\frac{it}{2}\sum_{\mu=1}^{2n}\gamma^\mu(\delta_\mu-\delta^{-1}_\mu)+m\gamma^{2n+1}
\nonumber\\
&+\frac{b}{2}\sum_{\mu=1}^{2n} \gamma^{2n+1}(\delta_\mu+\delta_\mu^{-1}-2),
\label{WilDir}
\end{alignat}
where $\gamma^\mu$ stands for the $\gamma$ matrix in $2n$ dimensions with 
$\{\gamma^\mu,\gamma^\nu\}=2\delta^{\mu\nu}$ as well as
$\gamma^{2n+1}=(-i)^n\gamma^1\cdots\gamma^{2n}$,
and   $\delta_\mu$ is the shift operator in the $\mu$ direction, defined by $\delta_\mu f_j=f_{j+\hat\mu}$.
\cin{In two dimensions, it is convenient to choose the $\gamma$ matrices as the Pauli matrices, 
$\gamma^a=\sigma^a$ ($a=1,2,3$),
while in four dimensions, one can choose $\gamma^a=\sigma^1\otimes\sigma^a$ ($a=1,2,3$), 
and $\gamma^{4,5}=\sigma^{2,3}\otimes\sigma^0$.}

First, we show in Fig. \ref{f:1ch_L}  the dependence of the computed $c_1$ 
on the truncation size $L$ in Eq. (\ref{c1Tru}).
When $L=0$ or $L=N$, $c_1$ vanishes. 
In the case of  $L=0$, this is because everything is truncated out.
On the other hand, 
for $L=N$, implying that nothing is truncated, $c_1$ also vanishes due to the presence of 
an artificial boundary at
$j_\mu=-N$ and $j_\mu=N-1$, which cancels with the contributions from the boundary at $j_\mu=0$.
Here, the term ``boundary" refers to the interface where the projector $\Pi_\mu$ transitions between 0 and 1.
Except for these trivial cases, the Kitaev formula yields nonzero values, although 
they are no longer exact integers, due to the artificial truncation.
Nevertheless, we find that choosing $L\sim N/2$ 
provides the most accurate results, allowing us to infer the expected Chern numbers.
This is quite reasonable, as a similar behavior is observed in the winding number as well \cite{PhysRevB.110.045437}.  
Therefore, we set \( L = N/2 \) in the following calculations.

\begin{figure}[htb]
\begin{center}
\begin{tabular}{c}
\includegraphics[width=0.95\linewidth]{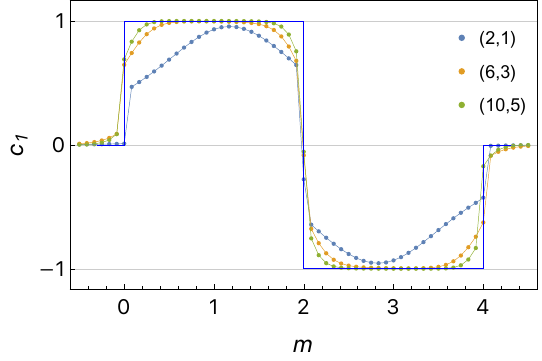}
\end{tabular}
\caption{
The first Chern number $c_1$ as a function of $ m$ ($t=b=1$) computed for the Wilson-Dirac model 
using the Kitaev formula.
The legends indicate the system size and truncation size, $(N,L)$.  
The blue lines represent the exact Chern numbers.
}
\label{f:1ch_m}
\end{center}
\end{figure}

In Fig. \ref{f:1ch_m}, the first Chern number is plotted as a function of $m$.  
A system size of $N=10$ (totaling $20^2$ sites) is sufficient to clearly distinguish the phases of the model.  
Remarkably, even with a minimal size $N=2$ ($4^2$ sites), the Kitaev formula qualitatively captures the phase diagram.  

\section{Second Chern number}\label{s:ch2}

So far, we have reviewed the real-space representation of the first Chern number derived by Kitaev.
The generalization to the second Chern number is straightforward.
In this section, we extend the Kitaev formula to the second Chern number 
and verify it by applying it to the Wilson-Dirac model.

\subsection{Kitaev formula in four dimensions}

From Eq. (\ref{GenChe}), the second Chern number is given by
\begin{alignat}1
c_2
&=\frac{1}{2!}\left(\frac{i}{2\pi}\right)^2\int d^4q\, 
\epsilon^{\mu\nu\rho\sigma}\tr P\partial_\mu P\partial_\nu P\partial_\rho P\partial_\sigma P.
\end{alignat}
The first step is to Fourier transform from momentum space to real space.
Following the same procedure as in Eq. (\ref{FouPos_1}), we obtain
\begin{alignat}1
&\epsilon^{\mu\nu\rho\sigma}\tr P[X_\mu,P][X_\nu,P][X_\rho,P][X_\sigma,P]_{ii}
\nonumber\\
&=\left(\frac{i}{2\pi}\right)^4\int d^4q\,\epsilon^{\mu\nu\rho\sigma}\tr P\partial_\mu P\partial_\nu P\partial_\rho P\partial_\sigma P.
\end{alignat}
The left-hand side does not depend on the site \(i\).  
The next step is to express the real-space position operator using the lattice projection operator  
defined in Eq. (\ref{LatPro}). Specifically,  
the four-dimensional counterpart of Eq. (\ref{ProPos_1}) is given by  
\begin{alignat}1
&\tr \epsilon^{\mu\nu\rho\sigma}P[X_\mu,P][X_\nu,P][X_\rho,P][X_\sigma,P]_{ii}
\nonumber\\
&=\Tr \epsilon^{\mu\nu\rho\sigma}P[\Pi_\mu,P][\Pi_\nu,P][\Pi_\rho,P][\Pi_\sigma,P]
\nonumber\\
&=\Tr \epsilon^{\mu\nu\rho\sigma}P\Pi_\mu P\Pi_\nu P \Pi_\rho P \Pi_\sigma P.
\end{alignat}
Thus, as a generalization of Eq. (\ref{1Che1}), we obtain the real space representation of the second Chern number,\begin{alignat}1
c_2
&=\frac{1}{2!}\left(\frac{2\pi}{i}\right)^2
\Tr \epsilon^{\mu\nu\rho\sigma}P\Pi_\mu P\Pi_\nu P \Pi_\rho P \Pi_\sigma P.
\label{SecChe}
\end{alignat}
This formula is sufficiently useful for calculating the second Chern number, the reason for which will be explained shortly.

In the case of the first Chern number, Eq. (\ref{1Che1}) can be simplified by eliminating overlapping regions and utilizing the deformation invariance of the partition into three regions.
Similarly, in the present case, such a simplification is also possible. Specifically, we divide the four-dimensional lattice into five generic regions labeled as $A=1,\dots,5$.
For a fixed set of five sites $i_1,\cdots,i_5$, we define
\begin{alignat}1
h_{i_1i_2i_3i_4i_5}&\equiv \epsilon^{abcde}\tr P_{i_ai_b}P_{i_bi_c}P_{i_ci_d}P_{i_di_e}P_{i_ei_a},
\label{DefNu2}
\end{alignat}
where $a,\dots,e$ take values from 1 to 5.
Then, similarly to Eq. (\ref{ThrReg}),
we have
\begin{alignat}1
\nu_2&\equiv
\sum_{i_1\in 1}\sum_{i_2\in 2}\sum_{i_3\in 3}\sum_{i_4\in 4}\sum_{i_5\in 5}
h_{i_1i_2i_3i_4i_5}
\nonumber\\
&=\Tr \epsilon^{ABCDE}P\Pi_A P\Pi_B P \Pi_C P \Pi_D P\Pi_E
\nonumber\\
&=\Tr \epsilon^{ABCD}P\Pi_A P\Pi_B P \Pi_C P \Pi_D P,
\label{Nu2}
\end{alignat}
where $\Pi_A$ is the lattice projector onto the region $A$, and 
in the last line, $A,B,C,D$ take values 1, 2, 3, and 4 only.
As an example of a generic five-partition of the four-dimensional lattice, we can define the following:
Starting with $\Pi_1=\Pi_x$, we take $\Pi_2=\Pi_y(1-\Pi_x)$,
$\Pi_3=\Pi_z(1-\Pi_x)(1-\Pi_y)$, $\Pi_4=\Pi_w(1-\Pi_x)(1-\Pi_y)(1-\Pi_z)$, and 
$\Pi_5$ the remaining region $\Pi_5=(1-\Pi_x)(1-\Pi_y)(1-\Pi_z)(1-\Pi_w)$.
This is one example of partitions of the four-dimensional lattice into five regions.
Other generic partitions included in (\ref{SecChe}) can be constructed, for example,
by starting with $\Pi_2=\Pi_y$, then successively removing the overlap regions,
$\Pi_1=\Pi_x(1-\Pi_y)$, and so on.
In this way, we obtain a total of $4!$ generic regions that contribute to Eq. (\ref{SecChe}).

Moreover, as in the case of the first Chern number, each contribution is the same due to the
invariance under deformation of the region. It is guaranteed by the conservation law
\begin{alignat}1
\sum_{i_5}h_{i_1i_2i_3i_4i_5}=0.
\end{alignat}
The invariance can be proved as follows:
The spectral projector satisfies $P^2=P$, which is explicitly expressed by $\sum_{k,l}P_{im,kl}P_{kl,jn}=P_{im,jn}$,
where $i,k,j$ stand for the sites while $m,l,n$ denote  the internal degrees of freedom. 
Using this property, we show that summing $ h_{i_1i_2i_3i_4i_5} $ in Eq. (\ref{DefNu2}) over one of $ i_\mu $, for instance, summing over $ i_5$ for all sites in the four-dimensional lattice, vanishes,
\begin{alignat}1
\sum_{i_5}h_{i_1i_2i_3i_4i_5}&=\tr \epsilon^{abcd}P_{i_ai_b}P_{i_bi_c}P_{i_ci_d}P_{i_di_a}
=0.
\label{Con2}
\end{alignat}
Now, let us focus our attention on a specific site $i_0\in5$, and define region $5'\equiv 5-\{i_0\}$,
where 5 stands for the set of sites in region 5. Then, 
\begin{alignat}1
\nu_2=h_{12345}=h_{12345'}+h_{1234i_0},
\end{alignat}
where we have used the shorthand notation, $h_{1i_2i_3i_4i_5}\equiv \sum_{i_1\in1}h_{i_1i_2i_3i_4i_5}$, and so on.
If the site $i_0$ is assigned to region 4, $\nu_2$ changes to
\begin{alignat}1
\nu'_2=h_{12345'}+h_{123i_05'}.
\end{alignat}
Their difference is 
\begin{alignat}1
\nu_2-\nu'_2=h_{1234i_0}-h_{123i_05'}.
\label{NuDif}
\end{alignat}
This  actually vanishes. To show this, note that 
the summation of Eq. (\ref{Con2}) over $i_1\in1$, $i_2\in2$, and $i_3\in3$ as well as the assignment $i_4=i_0$ yield
\begin{alignat}1
0=h_{123i_0(1+2+3+4+i_0+5')}
=h_{123i_04}+h_{123i_05'}.
\end{alignat}
This ensures that Eq. (\ref{NuDif}) indeed vanishes, telling that even 
if the boundary between the regions 4 and 5 are deformed,  $\nu_2$ is invariant.
Since $h_{i_1i_2i_3i_4i_5}$ is antisymmetric with respect to
its indices $i_1,\cdots,i_5$, it turns out  that $\nu_2$ is invariant 
under general deformation of the five regions.
Thus, we conclude that
\begin{alignat}1
c_2=\frac{4!}{2!}\left(\frac{2\pi}{i}\right)^2
\Tr \epsilon^{ABCDE}P\Pi_A P\Pi_B P \Pi_C P \Pi_D P\Pi_E.
\label{2ReaChe}
\end{alignat}
This is one of the main results of the paper, i.e., the real space representation of the second Chern number.
More generically in $2n$ dimensions, the $n$th Chern number can be expressed in real $2n$-dimensional space as
\begin{alignat}1
c_n=&\frac{(2n)!}{n!}\left(\frac{2\pi}{i}\right)^n
\nonumber\\
&\times\Tr \epsilon^{A_1A_2\cdots A_{2n}A_{2n+1}}
P\Pi_{A_1} P\Pi_{A_2} \cdots P\Pi_{A_{2n+1}}.
\label{nReaChe}
\end{alignat}
Equations (\ref{2ReaChe}) and (\ref{nReaChe}) are important as they demonstrate that higher Chern numbers can also be described using the Kitaev formula. 
However, even for the second Chern number, where calculations must be performed 
with a limited number of sites along each axis, the invariance under partitioning into five regions is slightly violated
due to the introduction of the truncation projector in the numerical calculations.
In such cases, the formula (\ref{SecChe}) is expected to provide more accurate results than Eq. (\ref{2ReaChe}).

\begin{figure}[htb]
\begin{center}
\begin{tabular}{c}
\includegraphics[width=0.99\linewidth]{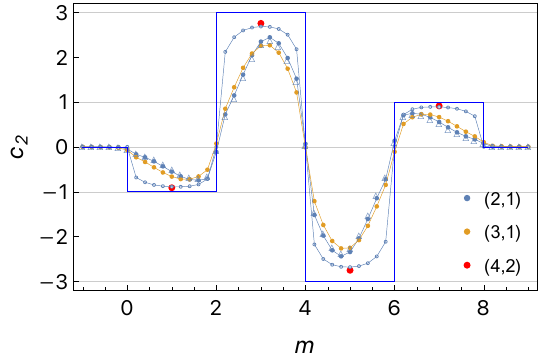}
\end{tabular}
\caption{
Second Chern number for the Wilson-Dirac model.
Colored connected points represent calculations using the real-space representation (\ref{SecChe}) with system and truncation sizes $(N,L) = (2,1)$, $(3,1)$, and $(4,2)$, as a function of $m$, where the other parameters are
set  $t=b=1$.
For the largest system, $(4,2)$, results are shown only for four red points at $m=1,3,5,7$ due to computational cost.
Open triangles indicate results for $(2,1)$ computed using Eq. (\ref{2ReaChe}).
Connected open small circles correspond to calculations performed in momentum space, where momentum space is discretized into a $12^4$ mesh.
The blue lines shows the exact second Chern numbers.
}
\label{f:2Che_p}
\end{center}
\end{figure}

\subsection{Application to the Wilson-Dirac model}

So far, we have derived the real-space representation of the second Chern number.  
To verify its validity, we apply it to the Wilson-Dirac model.

In Fig.~\ref{f:2Che_p}, we present the second Chern number computed using the real-space representation~(\ref{SecChe}) as a function of $m$.  
Compared to the first Chern number in two dimensions, the system sizes for the four-dimensional calculations are severely limited, ranging from $N=2$ (totaling $4^4$ sites) to $N=4$ ($8^4$ sites).  
For this reason, we first evaluate the second Chern number using Eq.~(\ref{SecChe}) rather than Eq. (\ref{2ReaChe}).
In particular, the calculation for the smallest system size, $(N, L) = (2,1)$, serves as a minimal test case.  
Despite its small size, the overall behavior of the second Chern number is qualitatively reproduced.  
For reference, we also show the second Chern number computed using Eq.~(\ref{2ReaChe}) 
for this minimal case, indicated by open triangles.  
As mentioned below Eq.~(\ref{nReaChe}), a small deviation from the values obtained via Eq.~(\ref{SecChe}) is observed.  
We have checked that this deviation decreases as the system size increases.  
Thus, in practical numerical calculations, Eq.~(\ref{2ReaChe}) can be used with sufficient accuracy.  

The system with $(N,L)=(3,1)$ is comparable to the system with $(2,1)$, despite its larger system size.  
This is probably because the truncation size $L$ of the former is not the optimal choice, $L = N/2$, as discussed in Sec.~\ref{s:2dwd}.  
Indeed, the system with $(4,2)$, which satisfies this condition, yields better results.  

Next, let us compare the results of the above real-space calculations with those in the conventional momentum space.
In the case of the first Chern number, the computational method of the U(1) Berry curvature
in a discretized momentum space has been established~\cite{FHS05}, 
which is manifestly gauge-invariant and gives strictly integer numbers.
However, in the case of the second Chern number, which inevitably involves non-Abelian Berry curvature, 
no rigorous algorithm for computation in momentum space has been developed so far.
Therefore, although the calculation is approximate, 
let us compare the momentum-space calculations~\cite{Mochol-Grzelak_2019} 
with the present real-space calculations.
In Fig.~\ref{f:2Che_p}, the momentum-space calculations with a $12^4$ mesh
are shown by small open circles. They yield results similar to those of the real-space calculations 
with the system size $(N,L)=(4,2)$. 
Although the computational cost of the momentum-space calculation is much lower than that of the real-space calculation,
the advantage of the latter is that it allows calculations in the presence of disorder in the model.

\begin{figure}[htb]
\begin{center}
\begin{tabular}{c}
\includegraphics[width=0.99\linewidth]{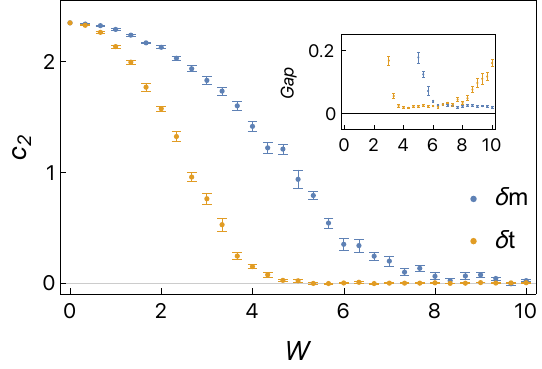}
\end{tabular}
\caption{
Ensemble average of the second Chern numbers for the disordered Wilson-Dirac model with sizes $(N,L)=(2,1)$.  
Orange dots indicate those of the model with random hopping parameters only, as a function of 
$W = W_t$ (with $W_m = 0$), whereas the blue dots show those with random mass parameters only, as a function of 
$W = W_m$ (with $W_t = 0$).  The parameters used are $m=3$, $t=b=1$.
Inset shows the averaged gap at half-filling. 
Each point shows an average over 10 ensembles.
}
\label{f:2Che_d}
\end{center}
\end{figure}

To investigate this further, we now examine a disordered Wilson-Dirac model.  
In Eq. (\ref{WilDir}), we define the model with a uniform kinetic term characterized by the parameter $t$,  
and a uniform mass term with the parameter $m$.  
We introduce disorder into these parameters, such that $t \to t + W_t \delta t_{j,\hat\mu}$,  
where $\delta t_{j,\hat\mu}$ represents an additional random hopping from site $j$ to site $j + \hat \mu$.  
Similarly, we introduce a random mass term $m \to m + W_m \delta m_j$,
where $\delta m_j$ is a random mass  on site $j$, for comparing the results  
in the case of disordered hoppings.  
We assume that $\delta t_{j,\hat\mu}$ is an independent random complex variable within the range of
$\delta t_{j,\hat\mu} \in [-1, 1]/2 + i[-1, 1]/2$, while $\delta m_j$ is an independent random real variable  
within the range of $\delta m_j \in [-1, 1]/2$.  
The parameters $W_t$ and $W_m$ control the strength of disorder.  

In Fig. \ref{f:2Che_d}, the second Chern numbers of the disordered Wilson-Dirac model are shown
as functions of the disorder strengths $W_t$ and $W_m$.
For the pure model, the second Chern number is $c_2 = 3$ (numerically, $c_2 \sim 2.35$ for this system size; see Fig. \ref{f:2Che_p}).
In the case of random hopping disorder, 
the half-filled gap closes as $W_t$ increases, and then reopens at larger $W_t$.
Accordingly, the second Chern number drops to zero, suggesting a topological transition from 
$c_2 = 3$ to $c_2 = 0$
without intermediate phases such as $c_2 = \pm1$.
Random mass disorder also appears to drive the $c_2 = 3$ phase to a trivial phase.
However, the gap remains nearly closed throughout, implying that the large random mass phase is not a trivial phase
but rather a gapless phase where the Chern number is ill-defined.

\section{Summary and Discussions}\label{s:sum}

The Kitaev formula for the first Chern number provides a topological description of the Chern number in real space. 
In addition to its conceptual significance, it is also quite useful for practical computations. 
In this paper, we first applied the Kitaev formula to the Wilson-Dirac model in two dimensions 
and explored its utility by testing the truncation scheme to identify the most suitable one.
We then generalized the Kitaev formula to the second Chern number and examined its validity. 
Despite the system size limitations for computations in higher-dimensional space, 
the second Chern number for the Wilson-Dirac model can be qualitatively reproduced. 
This opens up possibilities for investigating topological transitions driven by disorder.

In higher dimensions, the computational limitation of the system sizes is severe, 
so it may be desirable to invent hybrid formulas that are defined 
partially in real space and partially in momentum space, as a potential future direction. 
For example, the Thouless pump is defined in 1+1 dimensions, requiring both real space and momentum space, 
as the time variable can be regarded as a periodic momentum. 
Indeed, spatial disorder is random in real space but remains constant in time, making it suitable for treatment in such a hybrid space.
One possible implementation of such a hybrid approach has been demonstrated in Ref.~\cite{onaya2025formallyexactrealspacerepresentation}, where the Chern number is evaluated through the change of Berry phase, based on the relation $F = dA$, rather than by computing the Chern number directly.
It would be desirable to develop similar hybrid formulations for the second Chern number as well,
in order  to investigate three-dimensional pump phenomena.

The Wilson-Dirac model shows only two kinds of the first Chern numbers in two dimensions, while in four dimensions it shows four kinds of the second Chern numbers. It would be interesting to explore whether the model exhibits multistep topological transitions driven by disorder.

\acknowledgements
This work was supported in part by a Grant-in-Aid for Scientific Research 
	(Grant No. 22K03448) from the Japan Society for the Promotion of Science.
	
%


\end{document}